\documentclass[
twocolumn,
pra,showpacs,preprintnumbers,amsmath,amssymb]{revtex4}

\usepackage{times}
\usepackage{graphicx}
\usepackage{amssymb}
\usepackage{amsthm}
\usepackage{amsmath}
\usepackage{dsfont}
\usepackage{bm}
\usepackage{mathrsfs}
\usepackage{bbold}

\def\XXint#1#2#3{{\setbox0=\hbox{$#1{#2#3}{\int}$ }
\vcenter{\hbox{$#2#3$ }}\kern-.6\wd0}}

\begin{document}

\title{Cosmological horizons radiate}
\author{Ulf Leonhardt}
\affiliation{
\normalsize{
Department of Physics of Complex Systems,
Weizmann Institute of Science, Rehovot 7610001, Israel}
}
\date{\today}

\begin{abstract}
Gibbons and Hawking [Phys. Rev. D {\bf 15}, 2738 (1977)] have shown that the horizon of de Sitter space emits radiation in the same way as the event horizon of the black hole. But actual cosmological horizons are not event horizons, except in de Sitter space. Nevertheless, this paper proves Gibbons' and Hawking's radiation formula as an exact result for any flat space expanding with strictly positive Hubble parameter. The paper gives visual and intuitive insight into why this is the case. The paper also indicates how cosmological horizons are related to the dynamical Casimir effect, which makes experimental tests with laboratory analogues possible. 
\end{abstract}

\maketitle

\section{Introduction}

Horizons are known \cite{Brout} to turn quantum fluctuations into radiation. The event horizon \cite{JacobsonParentani} of a black hole radiates \cite{Hawking}. The causal horizon \cite{JacobsonParentani} of accelerated observers is predicted to radiate \cite{Unruheffect}. What about cosmological horizons \cite{Harrison}, will they radiate as well? Gibbons and Hawking have shown \cite{GibbonsHawking} that exponentially expanding flat space --- de Sitter space \cite{deSitter} --- emits thermal radiation with temperature 
\begin{equation}
k_\mathrm{B} T = \frac{\hbar H}{2\pi}
\label{gh}
\end{equation}
where $k_\mathrm{B}$ denotes Boltzmann's constant, $\hbar$ the reduced Planck constant and $H$ the Hubble parameter \cite{LL2}
\begin{equation}
H = \frac{\dot{a}}{a}
\label{hubble}
\end{equation}
that is constant for an expansion factor $a$ growing exponentially with cosmological time $t$. This regime of exponential growth is believed to be the asymptotic limit of the expanding universe. The actual expansion \cite{Riess2007} is non--exponential with varying Hubble parameter $H$ \cite{Weinberg,Confusion}. Will Gibbons' and Hawking's result (\ref{gh}) remain valid?

This question appears far from being trivial if we consider what cosmological horizons are and what they are not \cite{Confusion}. Imagine an arbitrary point co--moving with the universe. According to Hubble's law \cite{Harrison} the rest of the universe appears to withdraw from this point with a velocity that grows as $H\ell$ with proper distance $\ell$. At some distance $\ell_H$ the expansion velocity $H\ell_H$ reaches the speed of light $c$. The sphere with radius $\ell_H$ around the point we call the Hubble sphere and its surface the cosmological horizon \cite{Remark}. One would expect that no light from beyond that sphere would ever enter it (and reach the point in its center) but this is not true \cite{Confusion}. In fact, all galaxies with redshifts $z>1.6$ lie beyond our horizon, and yet they are visible. Not to mention the Cosmic Microwave Background that originates from $z\sim10^3$. Figure \ref{horizon} shows how this is possible. The figure also illustrates that the asymptotic de Sitter horizon does prevent light from entering its Hubble sphere. The de Sitter horizon is an event horizon \cite{JacobsonParentani} which justifies Gibbons' and Hawking's theory \cite{GibbonsHawking}, but the actual cosmological horizon is not. So will it radiate? And if it does what is its temperature?

The answer to this question is not entirely academic, despite the temperature of Eq.~(\ref{gh}) being in the range of $10^{-29}\mathrm{K}$ for the present Hubble parameter \cite{Planck,RiessMagellan,Holicow}. It turns out \cite{Annals} that the Gibbons--Hawking effect is essential for establishing the correct order of magnitude of the cosmological constant $\Lambda$ from the Lifshitz theory of vacuum fluctuations \cite{Rodriguez,Forces}. Furthermore, $\Lambda$ responds to changes in the inverse Gibbons--Hawking temperature with time, which may resolve \cite{Dror} one of the major puzzles in contemporary astrophysics \cite{Tension}, the tension between the present Hubble parameter inferred from the early \cite{Planck} or the late \cite{RiessMagellan,Holicow} cosmic evolution. The literature disagrees whether Eq.~(\ref{gh}) holds in general with arguments in favor \cite{JacobsonParentani,Annals,Aye} or against \cite{Nay}. Reference \cite{Annals} uses a fluid--mechanical analogue \cite{Analogues} to establish Eq.~(\ref{gh}) for arbitrary cosmic expansion. Here I prove Eq.~(\ref{gh}) without analogues as an exact result for any $H(t)>0$ (and Friedmann--Lema\^{i}tre--Robertson--Walker metric with zero spatial curvature).

\section{Visualization}

\begin{figure}[t]
\begin{center}
\includegraphics[width=20pc]{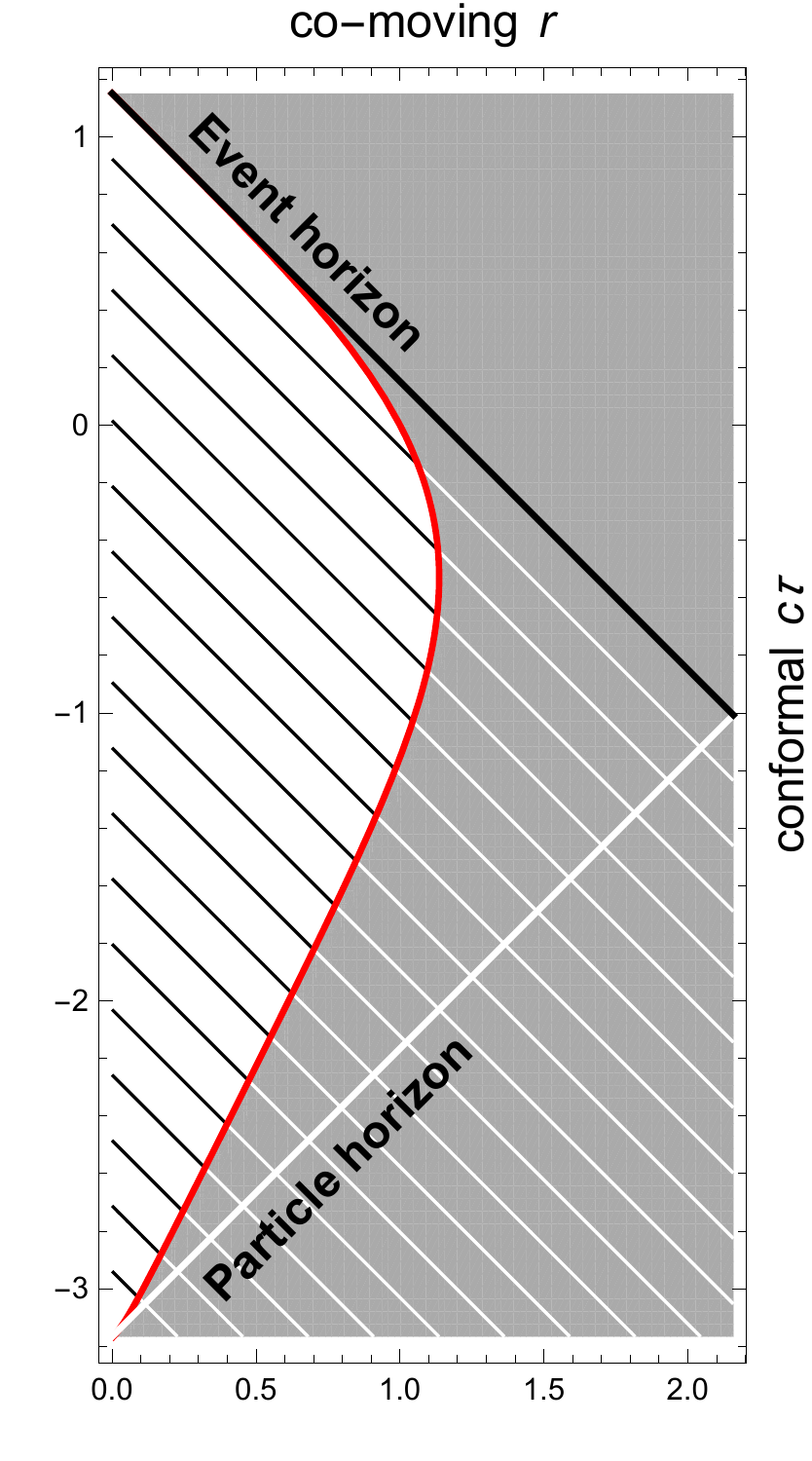}
\caption{
Horizons. Radial space--time diagram showing the entire history of the universe \cite{LambdaCDM} in (finite) conformal time $\tau$ and at co--moving radius $r$ from a point placed at the coordinate origin (length unit: horizon at present time \cite{LambdaCDM}). The cosmological horizon (red) separates the inside (white) from the outside (gray) of the Hubble sphere. Light propagates along the diagonals of the space--time diagram, as in Minkowski space, and freely crosses the cosmological horizon, unless the horizon becomes light--like as in the final stage of cosmic evolution when the universe approaches de Sitter space. In the early stage the universe is radiation--dominated and the cosmological horizon coincides with the particle horizon.
\label{horizon}}
\end{center}
\end{figure}

For simplicity, and in agreement with astronomical observation \cite{Curvature}, space is assumed to be flat with time--dependent length scale, the expansion factor $a(t)>0$. Space--time is not flat; the space--time metric reads 
\begin{equation}
ds^2 = c^2dt^2-a^2\,d\bm{r}^2
\label{ds2}
\end{equation}
and has non--zero Riemann curvature for non--zero $H$ \cite{CurvatureRemark}. In conformal time 
\begin{equation}
\tau = \int \frac{dt}{a} = \int \frac{d\alpha}{e^\alpha H} \quad \mathrm{with}\quad \alpha = \ln a
\label{tau}
\end{equation}
the metric (\ref{ds2}) reduces to $ds^2=a^2(c^2d\tau^2-d\bm{r}^2)$. Light rays ($ds=0$) thus propagate in conformal time $\tau$ and co--moving space $\bm{r}$ like in flat Minkowski space --- along the diagonals in a $(c\tau,\bm{r})$ space--time diagram (Fig.~\ref{horizon}).

Consider radially--incident rays. They encounter the cosmological horizon at the co--moving radius $r$ where, from the perspective of an observer at the origin, space appears to expand at the speed of light, {\it i.e.}\ where the proper distance $ar$ changes with $c$, which happens at $r=c/\dot{a} = c/(aH)$. Picture the horizon in a radial $(c\tau,r)$ space--time diagram (Fig.~\ref{horizon}). For the horizon to separate the stream of incident light, the horizon ($r/c=e^{-\alpha} H^{-1}$ with $\alpha=\ln a$) must be light--like as well ($r/c=\tau_0-\tau$ with $\tau_0=\mathrm{const}$). This is the case if the integrand of $\tau$ with respect to $\alpha$ [Eq.~(\ref{tau})] equals minus the integral, which requires the integrand $e^{-\alpha} H^{-1}$ to be proportional to $e^{-\alpha}$ and hence $H=\mathrm{const}$. Therefore, only in de Sitter space the cosmological horizon separates light into the inside and the outside of the Hubble sphere, only the de Sitter horizon is an event horizon. Otherwise light is able to cross the cosmological horizon. For light reaching us at present time with $a=1$ and $\tau_0=\left. \tau\right|_{\alpha=0}$ of Eq.~(\ref{tau}) we find for the parameters of the cosmological standard model \cite{LambdaCDM} that the crossing has occurred at redshift \cite{Weinberg} $z=e^{-\alpha}-1 \approx 1.6$ mentioned in Sec.~I. The light from galaxies with larger redshifts and the Cosmic Microwave Background reaches us from beyond the cosmological horizon, because the horizon is not light--like. 

Consider now the opposite situation: light or some other particles emitted from a point taken as the coordinate origin. The surface they can reach is called the particle horizon (Fig.~\ref{horizon}). The particle horizon ($r/c=\tau-\tau_1$ with $\tau_1=\mathrm{const}$) coincides with the cosmological horizon ($r/c=e^{-\alpha} H^{-1}$) if the integrand of $\tau$ with respect to $\alpha$ [Eq.~(\ref{tau})] equals the integral, which requires $H$ to be proportional to $e^{-2\alpha}=a^{-2}$. This is the equation of state of the radiation--dominated universe \cite{Weinberg}. So, fittingly, in the radiation--dominated era \cite{Weinberg} the particle horizon is also the cosmological horizon. 

\begin{figure}[t]
\begin{center}
\includegraphics[width=20pc]{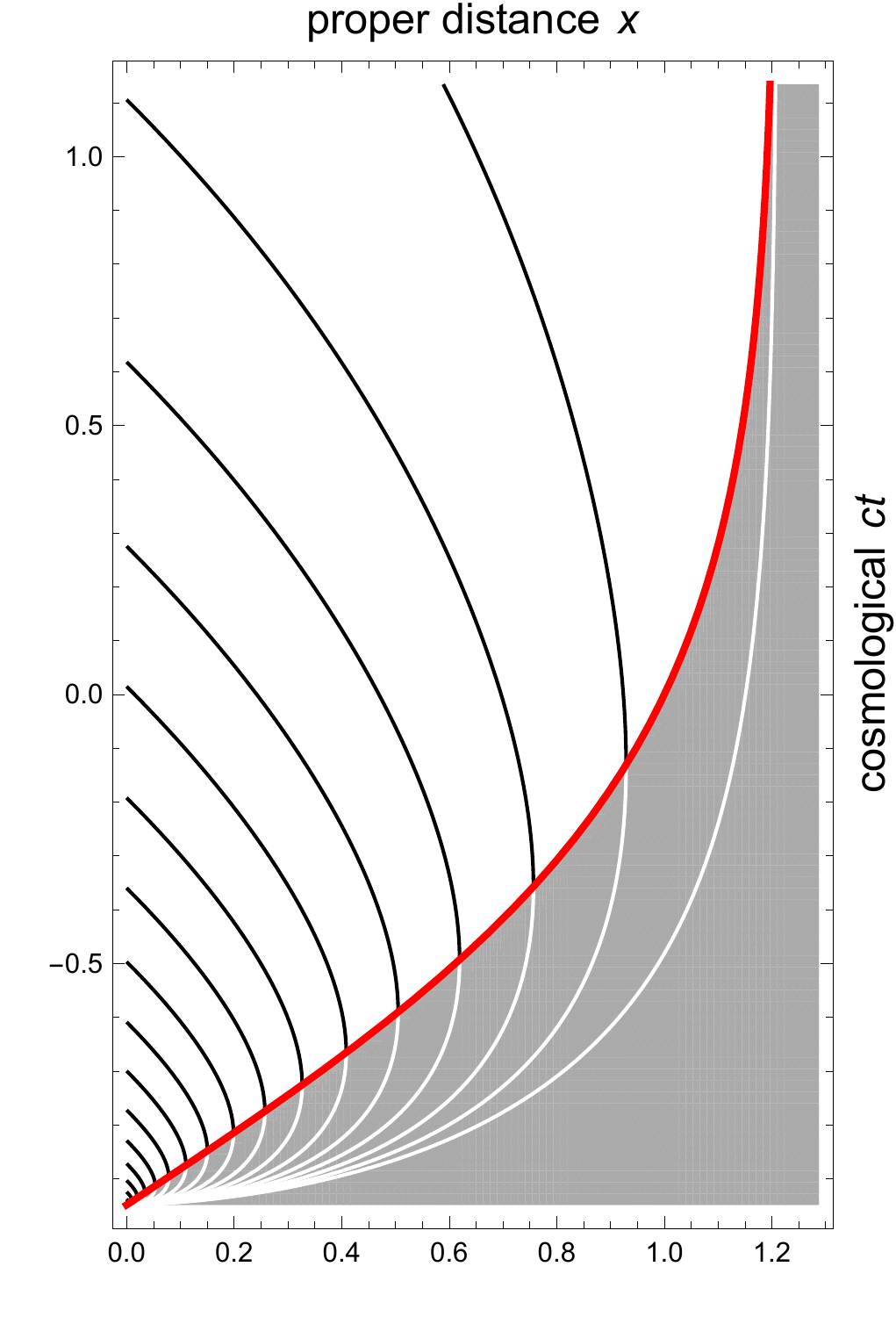}
\caption{
Moving medium. Viewed in the radial space--time diagram of cosmological time $t$ and proper distance $x$ the universe appears to light like a fluid expanding with velocity $Hx$ in agreement with Hubble's law (units as in Fig.~\ref{horizon}). At the cosmological horizon (red) the flow reaches the speed of light. The figure shows the same rays as in Fig.~\ref{horizon}. The rays are incident in the region (gray) outside the horizon, but dragged outward by the medium until the Hubble parameter $H$ has fallen such that they may enter the Hubble sphere (white). At the cosmological horizon the rays are vertical in the space--time diagram: light stands still, although only for a fleeting moment. In de Sitter space, when $H=\mathrm{const}$, this moment would last forever. 
\label{flow}}
\end{center}
\end{figure}

Figure \ref{horizon} illustrates the horizons in a space--time diagram of conformal time $\tau$ and co--moving radius $r$ where light propagates as in Minkowski space. Alternatively, one can also compensates for the spatial expansion factor $a$ in a diagram of cosmological time $t$ and proper distance $\ell=a r$ (Fig.~\ref{flow}). Consider the coordinates \cite{Annals}
\begin{equation}
\bm{x} = a \bm{r} \,.
\end{equation}
One obtains from the line element (\ref{ds2}) the metric 
\begin{equation}
ds^2=c^2dt^2 - (d\bm{x}-H\bm{x} \,dt)^2 \,.
\end{equation}
The null--geodesics in the $\{t,\bm{x}\}$ coordinates thus move with velocities 
\begin{equation}
\bm{v} = \frac{d\bm{x}}{dt} = \bm{c} + \bm{u}
\label{add}
\end{equation}
with $|\bm{c}|=c$ and 
\begin{equation}
\bm{u} = H \bm{x} \,.
\label{hubbleflow}
\end{equation}
Consequently, in $\{t,\bm{x}\}$ coordinates light propagates with $c$, but perceives the expanding universe as a moving medium \cite{Nonrelativistic} with outward flow $\bm{u}$ growing in speed $u$ with proper distance as $H\ell$ (Hubble's law) \cite{Annals}. At the cosmological horizon the medium reaches the speed of light. Figure \ref{flow} shows how incident light rays get to a momentary halt at the horizon. For exponentially expanding space, the proper distance to the horizon, $c/H$, remains constant, and so light is trapped there forever. For the actual expansion of the universe, however, the proper distance to the horizon increases --- the cosmological horizon appears to move outwards (Fig.~\ref{flow}). Light trapped for a fleeting moment there gets released and moves on. Nevertheless, this moment of light standing still will suffice for Gibbons--Hawking radiation. 

\begin{figure}[t]
\begin{center}
\includegraphics[width=20pc]{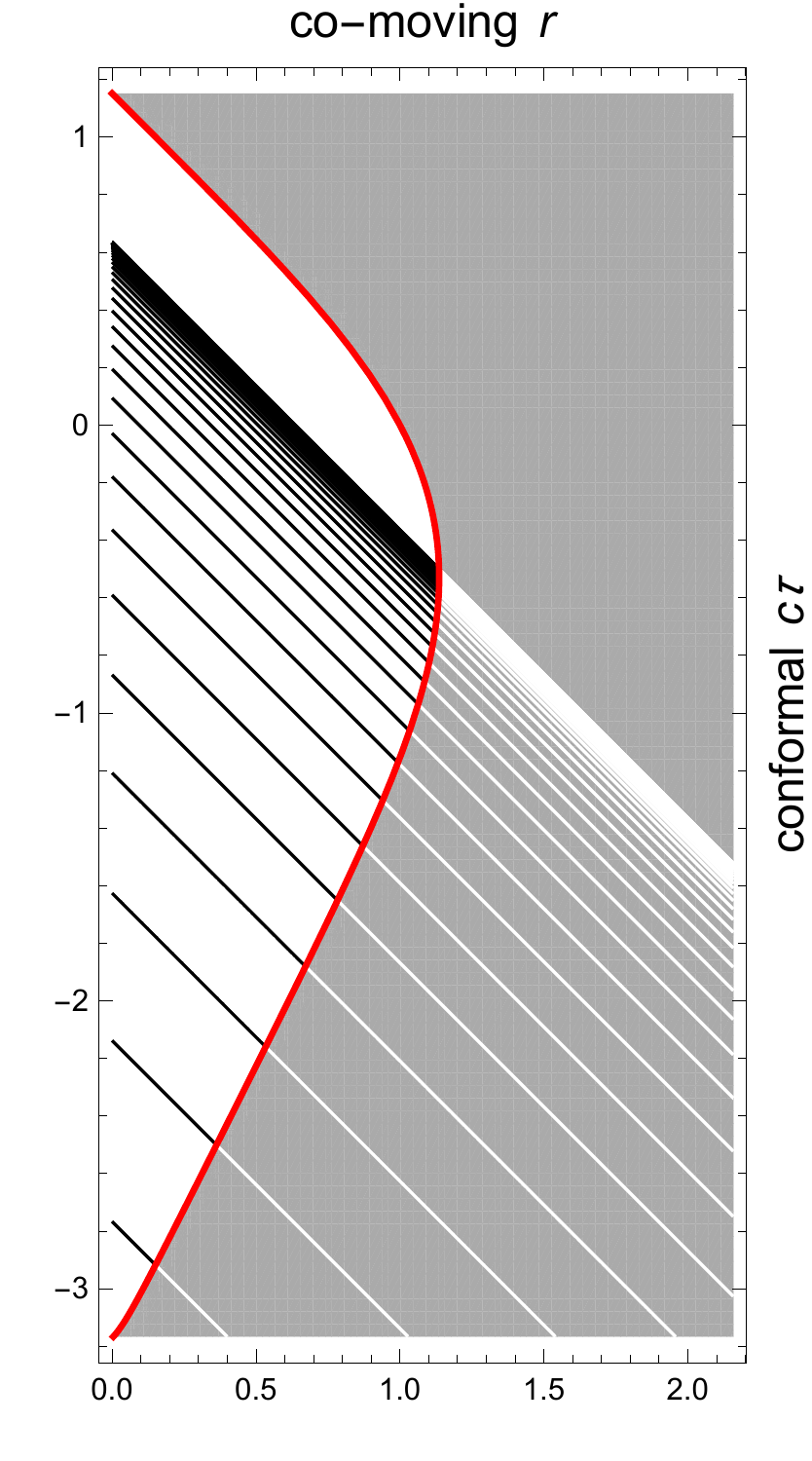}
\caption{
Wave of constant co--moving frequency. Contour lines of the phase [Eq.~(\ref{phi}) for $\tau-\tau_0+r/c<(a_0H_0)^{-1}$] shown in the space--time diagram of conformal time and co--moving radius for $\tau_0=-0.5$ (units as in Fig.~\ref{horizon}). The contour lines represent light rays incident from outside (gray) the Hubble sphere and crossing to the inside (white) at the cosmological horizon (red). For $\tau\ge \tau_0$ they are confined within the Hubble sphere. At $\tau=\tau_0+(a_0H_0)^{-1}$ the last ray reaches the origin. All incident radiation confined within the Hubble sphere at $\tau_0$ can be represented as superposition of these waves with different constant co--moving frequencies.
\label{wave}}
\end{center}
\end{figure}

To see this, consider a specific but arbitrary moment in time labelled by the conformal time $\tau_0$ with the expansion factor $a_0$ and Hubble parameter $H_0$. Imagine a light wave of constant frequency $\omega'$ in frames co--moving with the Hubble flow (\ref{hubbleflow}). Such waves may exist as mathematical objects, regardless whether one subscribes to their fluid--mechanical interpretation, and we may regard all incident light waves as superpositions of them. The co--moving frequency $\omega'$ is related to $\omega$, the frequency with respect to cosmological time, by the Doppler formula \cite{Doppler}
\begin{equation}
\omega' = \left(1-\frac{H|\bm{x}|}{c}\right) \omega \,.
\label{doppler}
\end{equation}
Expressed in terms of the co--moving radius $r$ we thus require
\begin{equation}
\left.\omega\right|_{\tau_0} = \frac{\omega'}{1-a_0H_0\, r /c} \,\,,\quad \omega'=\mathrm{const.}
\label{omega}
\end{equation}
The frequency $\omega$ is related to the phase $\varphi$ as $\omega=-\partial_t\varphi$. As the wave propagates in conformal time and co--moving radius as in Minkowski space (Fig.~\ref{horizon}) the phase of an incident radial wave must depend on $r+c\tau$. We thus obtain by integration
\begin{equation}
\varphi = \frac{\omega'}{H_0} \ln \left| 1 + a_0 H_0 \left(\tau_0-\tau - \frac{r}{c}\right)\right| . 
\label{phi}
\end{equation}
Consider now the wavenumber $k=\partial_r\varphi$. Inside the region with $r/c+\tau-\tau_0 < (a_0 H_0)^{-1}$ that coincides with the Hubble sphere at $\tau=\tau_0$ the wavenumber $k$ is negative, whereas outside this region $k>0$. Purely incident waves must have entirely negative spatial Fourier components and hence negative wavenumbers. Therefore we must require that the wave vanishes for $r/c+\tau-\tau_0 > (a_0 H_0)^{-1}$. This does not mean that the wave ends at the cosmological horizon throughout its history (Fig.~\ref{wave}). The edge of the wave coincides with the horizon only at $\tau=\tau_0$. As we can decompose all radially incident waves inside the temporary horizon at $\tau_0$ in terms of waves of constant but different co--moving frequencies $\omega'$, all waves reaching the origin are superpositions of waves with an edge. 

The quantum vacuum, however, is not confined by such edges, for the following simple reason \cite{Annals}: The universe is homogeneous and isotropic on cosmological scales \cite{Gott} and so must be the vacuum. One can move the coordinate origin to any spatial point, but the vacuum must remain invariant. This would be impossible if the quantum vacuum were confined by wave edges around a given point. From this follows that the radiation incident on any point must not be in the vacuum state. Cosmological horizons radiate. To work out the details takes a calculation. 

\section{Calculation}

\subsection{The model}

Consider a simple model for massless bosons such as the photons of the electromagnetic field: a conformally--coupled, massless scalar field \cite{BD} described by the mode decomposition
\begin{equation}
\hat{A}= \sum_{\bm{k}}\left( A_{\bm{k}}  \hat{a}_{\bm{k}} + A_{\bm{k}} ^* \hat{a}_{\bm{k}}^\dagger\right)
\end{equation}
where the mode functions $A_{\bm{k}}$ obey the wave equation \cite{BD}
\begin{equation}
\frac{1}{\sqrt{-g}}\partial_\mu \sqrt{-g}\, g^{\mu\nu} \partial_\nu A + \frac{R}{6} A = 0
\label{wave0}
\end{equation}
in terms of the metric tensor $g_{\alpha\beta}$, its determinant $g$ and inverse $g^{\mu\nu}$, while $R$ denotes the curvature scalar \cite{LL2}. The mode operators $\hat{a}$ and $\hat{a}^\dagger$ shall obey the Bose commutation relations. This simple model captures the essence of conformally--invariant bosonic fields experiencing the Gibbons--Hawking effect while avoiding technicalities associated with their polarization and internal structure. 

For the Bose commutation relation to hold for the mode operators, the mode functions are required \cite{LeoBook} to be orthonormal with respect to the scalar product \cite{BD}
\begin{equation}
\left(A_1,A_2\right) = \frac{i c}{\hbar} \int \left( A_1^* \,\partial^0 A_2 - A_2\, \partial^0 A_1^*\right) \sqrt{-g}\,dV 
\label{scalar}
\end{equation}
that, as a consequence of the wave equation (\ref{wave0}), does not depend on time. 

Specifically, the quantum field $\hat{A}$ shall evolve in a space--time geometry with metric (\ref{ds2}) expressed in spherical coordinates:
\begin{equation}
ds^2=c^2dt^2-a^2(dr^2+r^2 d\theta^2 + r^2\sin^2\theta\, d\phi^2)
\label{dsp}
\end{equation}
with curvature scalar \cite{LL2}
\begin{equation}
R= - \frac{6}{c^2}\,\left(\dot{H}+2H^2\right)
\label{r}
\end{equation}
and $\sqrt{-g}=a^3r^2\sin\theta$. We express the mode functions in terms of the spherical harmonics $Y_{lm}$ as
\begin{equation}
A = a(t)\,A_l(t,r)\,Y_{lm}(\theta,\phi)
\label{ylm}
\end{equation}
and arrive from Eqs.~(\ref{wave0}) and (\ref{dsp}-\ref{ylm}) at the equation of the partial waves $A_l$:
\begin{equation}
\partial_\tau^2 A_l = c^2\left(\partial_r^2+\frac{2}{r}\partial_r\right)A_l - \frac{c^2}{r^2}l(l+1)A_l
\label{weq}
\end{equation}
expressed in terms of conformal time $\tau$ defined in Eq.~(\ref{tau}). In flat Minkowski space one gets exactly the same wave equation. So not only do null geodesics propagate in conformal time and co--moving space as in Minkowski space, the full quantum field $\hat{A}$ does it as well. This is a consequence of its conformal invariance \cite{BD}. For guidance and intuition, consider first the case $l=0$ of purely radial propagation, before proceeding to the general $l\ge 0$. 

\subsection{Radial propagation}

Put $l=0$. In this case the general solution of the radial wave equation (\ref{weq}) takes the familiar d'Alembert form 
\begin{equation}
A_0 = \frac{1}{r}\left[f_+(r+c\tau)+f_-(r-c\tau)\right]
\end{equation}
with the functions $f_\pm$ describing incoming (+) and outgoing (-) waves. Consider in particular
\begin{equation}
A_0 = \frac{\cal A}{r}\left(\eta\mp\rho\right)^{i\nu+1}
\label{a0}
\end{equation}
with constant ${\cal A}$, the dimensionless variables
\begin{equation}
\eta = 1 + H_0 a_0 (\tau_0-\tau) \,,\quad \rho = \frac{H_0 a_0}{c}\, r
\label{etarho}
\end{equation}
and the dimensionless constant
\begin{equation}
\nu = \frac{\omega'}{H_0} \,.
\label{nu}
\end{equation}
The radial wave given by Eqs.~(\ref{a0}-\ref{nu}) with the minus sign in Eq.~(\ref{a0}) has the phase profile (\ref{phi}) of a wave with constant co--moving frequency discussed in Sec.~II and illustrated in Fig.~\ref{wave}. There we argued that for being strictly ingoing the wave needs to be confined in the region $\rho<\eta$ (Fig.~\ref{wave}). 

After the incident wave focuses at the origin it is reflected there and leaves as an outgoing wave. We thus write for the propagation of the incident mode:
\begin{equation}
A_+ = \frac{\cal A}{r}\left(\Theta(\eta-\rho)\left(\eta-\rho\right)^{i\nu+1} - \left(\eta+\rho\right)^{i\nu+1} \right) .
\label{aplus}
\end{equation}
The minus sign between the incoming and the outgoing term ensures that $A_+$ does not diverge at the origin. Instead, it forms there a diffraction--limited spot \cite{Fink}. There is one more subtlety to consider: Eq.~(\ref{aplus}) holds only for $\eta\ge 0$. At $\eta=0$ the incident wave runs out and vanishes for $\eta<0$ (Fig.~\ref{wave}). No further light can by reflected; the edge of the last reflected wave moves as $\rho=-\eta$ with falling $\eta$ given by Eq.~(\ref{etarho}). We indicate this by writing a $\Theta(\eta+\rho)$ in front of the reflected wave  for $\eta<0$:
\begin{equation}
A_+ = -\frac{\cal A}{r}\,\Theta(\eta+\rho)\left(\eta+\rho\right)^{i\nu+1} .
\label{aplusm}
\end{equation}

\begin{figure}[t]
\begin{center}
\includegraphics[width=20pc]{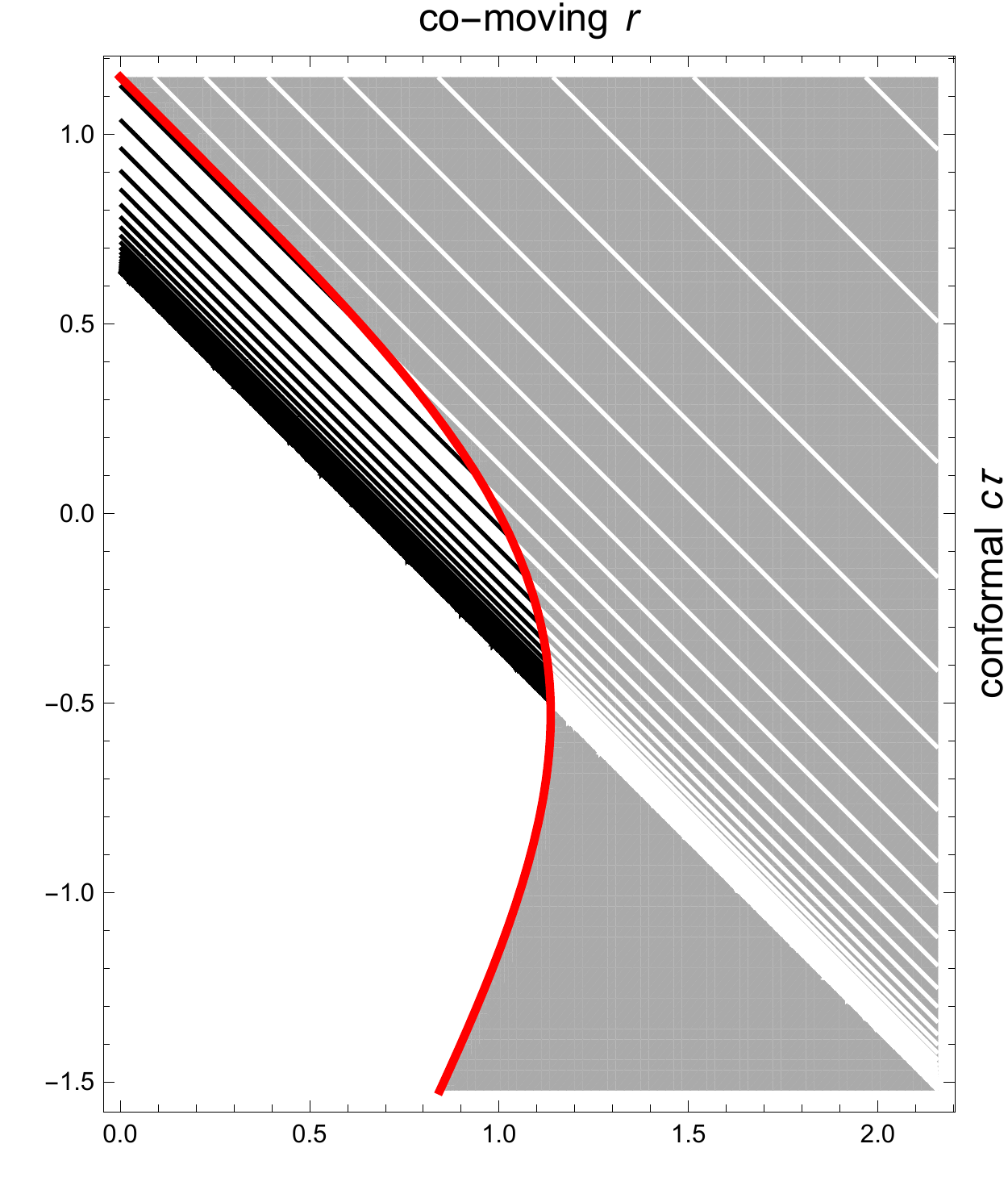}
\caption{
Partner of the wave with constant co--moving frequency shown in Fig.~\ref{wave}. Contour lines of the phase [Eq.~(\ref{phi}) for $\tau-\tau_0+r/c>(a_0H_0)^{-1}$] with units as in Fig.~\ref{wave}. As there the contour lines represent light rays. One sees that the partner wave reaches the origin at $\tau>\tau_0+(a_0H_0)^{-1}$. The quantum vacuum entangles the partners in an Einstein--Podolski--Rosen state. Reduced to one of the partners, this state appears as thermal radiation with Gibbons--Hawking temperature (\ref{gh}). 
\label{partner}}
\end{center}
\end{figure}

Now imagine a wave $A_-$ of constant co--moving frequency that stays outside $\rho>\eta$ (Fig.~\ref{partner}). Here, according to Eq.~(\ref{omega}), the co--moving frequency $\omega'$ has the opposite sign of the frequency with respect to cosmological time, $\omega$. For having a wave oscillating with positive frequencies $\omega$ the co--moving frequency needs to be negative, and so does the parameter $\nu$ in Eq.~(\ref{nu}). We thus complex--conjugate the solution of Eq.~(\ref{a0}) and define $A_-$ as
\begin{equation}
A_- = \frac{\cal A}{r}\,\Theta(\rho-\eta)\left(\rho-\eta\right)^{-i\nu+1} .
\label{aminus}
\end{equation}
As Eq.~(\ref{aplus}) this equation is only valid for $\eta>0$ and hence $\tau-\tau_0<(a_0H_0)^{-1}$. At later times $A_-$ reaches the origin and is reflected there:
\begin{equation}
A_- = \frac{\cal A}{r}\left((\rho-\eta)^{-i\nu+1}-\Theta(-\eta-\rho)\left(-\eta-\rho\right)^{-i\nu+1} \right) 
\label{aminusm}
\end{equation}
for $\eta<0$. The need for complex conjugation in the definition of $A_-$ also becomes evident from normalizing $A_-$ with respect to the scalar product defined in Eq.~(\ref{scalar}). We recall that the scalar product does not depend on time and evaluate it at $\eta=0$ where we get
\begin{equation}
\left(A_1,A_2\right) = \frac{8\pi {\cal A}^2 H_0 \nu}{\hbar} \int_0^\infty \rho^{i(\nu_1-\nu_2)+1} d\rho
\end{equation}
for $\nu_1\sim\nu_2\sim\nu$ (the scalar product vanishes if $\nu_1\neq\nu_2$ for otherwise it would be time--dependent). We substitute $\exp(-\xi)$ for $\rho$ and obtain the standard integral of the delta function:
\begin{equation}
\left(A_1,A_2\right) = \frac{(4\pi)^2 {\cal A}^2 H_0 \nu}{\hbar} \, \delta(\nu_1-\nu_2) \,.
\end{equation}
The norm is positive, as required, and it were negative without the complex conjugation in the definition of $A_-$. For $A_+$ we also evaluate the scalar product at $\eta=0$ and obtain exactly the same norm. 

We have thus established two sets of modes with parameter $\nu$, one ($A_+$) describing radiation incident from inside $\rho<\eta$ and the other ($A_-$) incident radiation staying outside. Similar to the closely--related Rindler modes \cite{Brout} these modes form a complete orthonormal set for incident radiation. 

Next we follow Damour's and Ruffini's elegant argument \cite{DR} and note that the quantum vacuum occupies modes that are analytic in $\rho-\eta$ (or $\rho+\eta$). In particular, they must not vanish for $\rho<\eta$ or $\rho>\eta$. It is easy to construct analytic $A_{\mathrm{vac}\pm}$ from the $A_\pm$ of Eqs.~(\ref{aplus}) and (\ref{aminus}):
\begin{equation}
A_{\mathrm{vac}\pm} = A_\pm \cosh\zeta + A_\mp^* \sinh\zeta 
\label{bogo}
\end{equation}
with 
\begin{equation}
\tanh\zeta = e^{-\pi\nu} \,.
\end{equation}
The Bogoliubov transformation (\ref{bogo}) preserves the orthonormality of the mode functions and the Bose commutation relations of the associated mode operators \cite{LeoBook}
\begin{equation}
\hat{a}_{\mathrm{vac}\pm} = \hat{a}_\pm \cosh\zeta - \hat{a}_\mp^\dagger \sinh\zeta \,.
\end{equation}
Defining the vacuum state $|0\rangle$ as the eigenstate of the $\hat{a}_{\mathrm{vac}\pm}$ with eigenvalue zero gives \cite{LeoBook}
\begin{equation}
|0\rangle = \frac{1}{\sqrt{Z}} \sum_{n=0}^\infty e^{-n\pi\nu} |n\rangle_+|n\rangle_-
\label{epr}
\end{equation}
in terms of the particle--number states $|n\rangle_\pm$ of the modes $A_\pm$ and with $Z^{-1}=1-e^{-2\pi\nu}$. The state described in Eq.~(\ref{epr}) is an Einstein--Podolski--Rosen state (a two--mode squeezed vacuum) \cite{LeoBook}, the strongest entangled state with given mean energy \cite{Barnett}. The $A_\pm$ modes are thus strongly correlated: a particle of frequency $\omega'$ detected at $\tau_0$ at the origin is accompanied by a partner particle in mode $A_-$. The partner may also appear at the origin after the $A_-$ mode comes in when the conformal time $\tau_0 + (a_0 H_0)^{-1}$ has elapsed, which may take a rather long cosmological time. 

The reduced quantum state for the $A_+$ mode is a thermal state \cite{LeoBook} with temperature $T$ given by 
\begin{equation}
2\pi\nu= \frac{\hbar\omega}{k_\mathrm{K} T}
\end{equation}
From definition (\ref{nu}) and Eq.~(\ref{omega}) at $r=0$ follows Gibbons' and Hawking's formula (\ref{gh}) for any $H_0$ and hence for any $H(t)>0$. 

\subsection{General propagation}

Radial propagation is sufficient for describing the field captured by a single point (with coordinate origin set to this point). Yet the field energy at this point depends on derivatives of the field amplitudes and hence on multipole momenta, and so do other correlation functions. It is therefore necessary to consider the general case of arbitrary angular momentum $l$. Inspired by the radial case of $l=0$, we seek the solution of the wave equation (\ref{weq}) as
\begin{equation}
A_l = \frac{{\cal A}_l}{r}\, (\eta\mp\rho)^{i\nu+1} p_l(\pm z)
\label{sol}
\end{equation}
with 
\begin{equation}
z= - \frac{\eta}{\rho}
\end{equation}
and find 
\begin{equation}
p_l(z)= \frac{(i\nu)!}{(i\nu-l)!}\,\,{}_2 F_1\left(-l,l+1,-i\nu,\frac{1-z}{2}\right)
\label{pl}
\end{equation}
in terms of Gauss' hypergeometric function \cite{Erdelyi}. The prefactor is chosen for later convenience. From the hypergeometric series follows that the $P_l(z)$ are polynomials of order $l$ in $z$, with the first two given by
\begin{equation}
p_0(z)=1 \,,\quad p_1(z) = 1+i\nu-z \,.
\label{pinit}
\end{equation}
We use Eq.~3.4.(6) of Ref. \cite{Erdelyi} to express Eq.~(\ref{pl}) in terms of Legendre functions and deduce from Eq.~3.8.(12) of Ref.~\cite{Erdelyi} the recurrence relation
\begin{equation}
p_{l+2}(z) + (2l+3) z\, p_{l+1}(z) +(l-i\nu)(l+2+i\nu)\,p_l(z) \,.
\label{recurrence}
\end{equation}
Applying this relation with the initial values of Eq.~(\ref{pinit}) we can easily compute the $p_l(z)$. In particular, we obtain for the constant term in the polynomials
\begin{equation}
p_l(0) = \prod_{m=1}^l (2m-l+i\nu) \quad\mathrm{for}\quad l>0 
\label{pl0}
\end{equation}
and $p_0=1$. These are all the mathematical preparations needed for constructing and normalizing the mode functions of the incident radiation. 

In analogy to Eqs.~(\ref{aplus}) and (\ref{aplusm}) we write for the mode incident inside $\rho<\eta$ the compact expression 
\begin{eqnarray}
A_+ &=& \frac{{\cal A}_l}{r}\left(\Theta(\eta-\rho)\left(\eta-\rho\right)^{i\nu+1} p_l(z) \right.
\nonumber\\
&& \left.  - (-1)^l\, \Theta(\eta+\rho) \left(\eta+\rho\right)^{i\nu+1} p_l(-z) \right) .
\end{eqnarray}
The incoming wave is reflected with coefficient $(-1)^l$ for the following reason. The highest singularity in $p_l(z)$ for  $r\rightarrow 0$ is given by the highest term in $z=-\eta/\rho$, which is proportional to $z^l$ in the $l$--th order polynomial $p_l(z)$. Subtracting the ingoing and outgoing term with the difference $(-1)^l$ thus removes the leading singularity. It also removes all other singularities, as there must exist a regular solution as linear combination of the two fundamental solutions (\ref{sol}). The so--constructed mode function describes light that reaches the origin at $\eta>0$ when $\tau-\tau_0<(a_0H_0)^{-1}$.

The light outside the cosmological horizon at $\tau_0$ propagates inwards as well. but reaches the origin at $\tau-\tau_0>(a_0H_0)^{-1}$ when $\eta<0$. We describe the corresponding mode function as 
\begin{eqnarray}
A_- &=& \frac{{\cal A}_l}{r}\left(\Theta(-\eta+\rho)\left(\eta+\rho\right)^{-i\nu+1} p_l(z) \right.
\nonumber\\
&& \left.  - (-1)^l\, \Theta(-\eta-\rho) \left(-\eta-\rho\right)^{-i\nu+1} p_l(-z) \right) \quad
\end{eqnarray}
where we took the complex conjugates of the fundamental solutions (\ref{sol}) for having a positive norm. The normalization is best done at the time when $\eta=0$ and hence $z=0$. We obtain for both $A_+$ and $A_-$:
\begin{equation}
\left(A_1,A_2\right) = \frac{(4\pi)^2 {{\cal A}_l}^2 |p_l(0)|^2\,H_0 \nu}{\hbar}\, \delta(\nu_1-\nu_2)
\end{equation}
with the $p_l(0)$ given by Eq.~(\ref{pl0}). We can thus proceed exactly as in Sec.~IIIB (without the need to consider gray--body factors \cite{Brout} as for the Hawking radiation of black holes). We obtain also for general wave propagation the Einstein--Podolski--Rosen state of Eq.~(\ref{epr}) as the vacuum state seen by the observer modes $A_\pm$. An observer co--moving with the universe would perceive the vacuum as a thermal state with Gibbons--Hawking temperature (\ref{gh}).

\section{Outlook}

Cosmological horizons radiate with Gibbons--Hawking temperature (\ref{gh}) in a spatially--flat expanding universe with $H(t)>0$. This paper has proven this statement as an exact result for a Friedmann--Lema\^{i}tre--Robertson--Walker metric with zero spatial curvature \cite{Weinberg}. We may also imagine a contracting universe with $H(t)<0$ where the Hubble flow of Eq.~(\ref{hubbleflow}) points inwards and grows in magnitude with growing proper distance. Also in this case a cosmological horizon is established when the Hubble flow reaches the speed of light. We can run the entire argument of the paper with $H$ replaced by $|H|$. and arrive at Eq.~(\ref{gh}) with $|H|$. An interesting --- and different --- scenario occurs when $H$ changes sign, in particular when $H$ is oscillating (as in anti de Sitter space \cite{Kolo}). In this case light may cross the cosmological horizon multiple times. Similar to black--hole lasing \cite{BHlaser} each interaction with the horizon may create radiation, but it depends on the phase acquired between the interactions whether the radiation is amplified or de--amplified. 

While the ``oscillating universe'' is a purely theoretical case in astrophysics, it is in fact the most realistic case for laboratory analogues of cosmological horizons and their radiation, complementing and generalizing the analogue of de Sitter space with a Bose--Einstein condensate \cite{FF}. The reason is that one can create periodic modulations of the refractive index \cite{Dezael,Wilson,Hakonen,Veccoli} that act like a periodically--modulated expansion factor in the space--time metric (\ref{ds2}). The Hubble constant is proportional to the modulation frequency $\omega_0$. When $\omega_0$ is comparable with the radiation frequency $\omega$ the effective temperature (\ref{gh}) becomes significant and detectable radiation is generated. This process is closely related to the dynamical Casimir effect \cite{Wilson,Hakonen,Veccoli} where a boundary or the optical length to a boundary is modulated. The radiation of de Sitter space has been mapped to the radiation produced  by an accelerated mirror \cite{Good}. The Gibbons--Hawking radiation of cosmological horizons may be regarded as a pure and intriguing case within the wider area of the dynamical Casimir effect \cite{DCE}.

\section*{Acknowledgements}

This paper is dedicated to the memory of Renaud Parentani. 
I am grateful to him for introducing me to the quantum physics of horizons, 
and I also thank
D. Bermudez,
D. Berechya,
and
N. Ebel for discussions directly related to this paper. 
The paper has been supported by the Israel Science Foundation and the Murray B. Koffler Professorial Chair. 



\begin{thebibliography}{99}
\bibitem{Remark}
Frequently the cosmological horizon is called the Hubble sphere, although it is not a sphere but the surface of a sphere.
\bibitem{CurvatureRemark}
From Eq.~(\ref{r}) follows that that the curvature scalar $R\neq 0$ unless $H=0$ (empty space) or $\partial_t H^{-1}=2$ (radiation--dominated universe). But even in the latter case the energy--momentum tensor and hence \cite{LL2} the Ricci tensor does not vanish.
\bibitem{LambdaCDM}
We use the standard $\Lambda$CDM model with parameters inferred from measurements of the Cosmic Microwave Background: $H^2=H_0^2(\Omega_\mathrm{R} a^{-4} + \Omega_\mathrm{M} a^{-3} + \Omega_\Lambda)$ with time in units of $1/H_0$ and $\Omega_\mathrm{R}=0.925\times10^{-4}$, $\Omega_\mathrm{M}=0.3153 \pm 0.0073$, $\Omega_\Lambda=0.6847 \pm 0.0073$.
\bibitem{Nonrelativistic}
Note that the addition theorem of velocities in Eq.~(\ref{add}) is non--relativistic, which is typical for analogues of gravity \cite{Analogues}.
\end{thebibliography}
\end{document}